\documentclass[twocolumn]{aastex62}

\include{definitions} 
\received{\today}

\shorttitle{A black hole in MAXI~J1820+070}
\shortauthors{Torres, M. A. P. et al. }

\begin{document}

\title{DYNAMICAL CONFIRMATION OF A  BLACK HOLE IN  MAXI~J1820+070}


\author[0000-0002-3348-4035]{M. A. P. Torres}
\affiliation{Instituto de Astrof\'isica de Canarias, 38205 La Laguna, Tenerife, Spain}
\affiliation{Departamento de Astrof\'\i{}sica, Universidad de La Laguna, E-38206 La Laguna, Tenerife, Spain}
\affiliation{SRON, Netherlands Institute for Space Research, Sorbonnelaan 2, NL-3584 CA Utrecht, the Netherlands}

\author{J.~Casares}
\affiliation{Instituto de Astrof\'isica de Canarias, 38205 La Laguna, Tenerife, Spain}
\affiliation{Departamento de Astrof\'\i{}sica, Universidad de La Laguna, E-38206 La Laguna, Tenerife, Spain}
\author{F. Jim\'enez-Ibarra}
\affiliation{Instituto de Astrof\'isica de Canarias, 38205 La Laguna, Tenerife, Spain}
\affiliation{Departamento de Astrof\'\i{}sica, Universidad de La Laguna, E-38206 La Laguna, Tenerife, Spain}
\author{T.~Mu\~noz-Darias}
\affiliation{Instituto de Astrof\'isica de Canarias, 38205 La Laguna, Tenerife, Spain}
\affiliation{Departamento de Astrof\'\i{}sica, Universidad de La Laguna, E-38206 La Laguna, Tenerife, Spain}

\author{M. Armas Padilla}
\affiliation{Instituto de Astrof\'isica de Canarias, E-38205 La Laguna, Tenerife, Spain}
\affiliation{Departamento de Astrof\'\i{}sica, Universidad de La Laguna, E-38206 La Laguna, Tenerife, Spain}

\author{P.G.~Jonker}
\affil{SRON Netherlands Institute for Space Research\\
Sorbonnelaan 2, 3584 CA Utrecht, The Netherlands}
\affil{Department of Astrophysics/IMAPP, Radboud University\\
P.O.~Box 9010, 6500 GL Nijmegen, The Netherlands}

\author{M.~Heida}
\affil{Cahill Center for Astronomy and Astrophysics, California Institute of Technology \\
1200 California Blv., Pasadena, CA 91125, USA}

\begin{abstract}
We present time-resolved 10.4-m GTC and 4.2-m WHT intermediate
resolution spectroscopy of the X-ray transient MAXI J1820+070 (=ASASSN-18ey) obtained during its
decline to the quiescent state. Cross-correlation of the 21 individual spectra against
late-type templates reveals a sinusoidal velocity modulation with a period of
$0.68549  \pm 0.00001 $ d and semi-amplitude of $417.7  \pm 3.9$ km
s$^{-1}$. We derive a mass function
f(M)=$5.18 \pm 0.15$ M$_\odot$, dynamically confirming the black hole nature of
the compact object. Our analysis of the stellar absorption
features supports 
a  K$3-5$ spectral classification for the donor star, which
contributes $\approx 20 \%$ of the total flux at 5200-6800 \AA. The
photometric $0.703 \pm 0.003$ d periodicity observed during outburst is 2.6 \% 
longer than the orbital period supporting the presence of a superhump
modulation in the outburst light curves. In line with this
interpretation, we constrain the binary mass ratio to be $q\simeq
0.12$. In addition, we observe a sharp increase in the H$\alpha$
emission line equivalent width during inferior conjunction of the  donor star that we interpret as a grazing eclipse of the accretion disc 
and allows us to constrain the binary inclination to
$i\gtrsim69^{\circ}$. On the other hand, the absence of X-ray eclipses during outburst imply $i\lesssim77^{\circ}$. These inclination limits, together with our 
dynamical solution, lead to  a black hole mass in the range 7--8 M$_\odot$.
We also measure a systemic velocity $\gamma=-21.6 \pm 2.3 $ km s$^{-1}$
which, combined with the Gaia DR2 proper motion and parallax, implies a large peculiar velocity $\sim$100 km s$^{-1}$.  
\end{abstract}

\keywords{accretion, accretion discs -- X-rays: binaries -- stars: black
 holes - stars: individual (MAXI J1820+070)}

\section{Introduction}\label{sec:intro}

Galactic black-hole (BH)  X-ray binaries provide an observational way
to study the formation of  these compact stellar remnants. In this
respect, dynamical studies serve to define the  BH stellar-mass
distribution (e. g. \citealt{1998ApJ...499..367B,2010ApJ...725.1918O,2011ApJ...741..103F,2012ApJ...757...36K}) that can be compared with those predicted from different supernova models
(e.g. \citealt{2001ApJ...554..548F}; \citealt{2012ApJ...757...91B}).
Further constraints on the BH formation mechanism can be
achieved from X-ray binaries with known space velocities.  Combined
with dynamical studies, these permit to search for
potential kicks suffered by BHs at formation, which provides
input to both supernova and binary evolution models
(\citealt{2004MNRAS.354..355J}; \citealt{2014PASA...31...16M};
\citealt{2015MNRAS.453.3341R}). Currently, the number of BH X-ray
binaries with a dynamical mass measurement is $18$
(\citealt{2014SSRv..183..223C}; \citealt{2016A&A...587A..61C,2016ApJS..222...15T}) 
while sources with constrained space velocity are more limited
(e.g. \citealt{2017hsn..book.1499C}; Atri et al. 2019, submitted).

The object of the dynamical study presented in this letter, MAXI
J1820+070 (hereafter J1820), is a BH candidate with a proper motion and parallax distance determination from GAIA DR2
(e.g.~\citealt{2019MNRAS.485.2642G}). J1820 was discovered on 6 March 2018 as
an optical transient by the All-Sky Automated Survey for Supernovae
(and named ASASSN-18ey; \citealt{2018ApJ...867L...9T}) and, as an X-ray
transient, by the Monitor of All-sky X-ray Image (MAXI;
\citealt{2009PASJ...61..999M}). With both an optical magnitude of
$g\sim$ 11.2 and an X-ray flux of $\sim$ 4 Crab
\citep{2009PASJ...61..999M} at outburst peak,  J1820  is among the
brightest X-ray transients ever observed. The optical
spectra displayed broad emission lines, characteristic of low-mass
X-ray binaries in outburst (\citealt{2018ApJ...867L...9T}), 
with strong and variable contribution from a disc outflow component 
 (\citealt{2019arXiv190604835M}). The system was  classified as a BH
 candidate based on its multi-wavelength properties \citep{2018ATel11399....1K,
 2018ATel11403....1K, 2018ATel11418....1B, 2019ApJ...874..183S,
 2018ApJ...867L...9T}.  A  tentative orbital period 
of $16.87 \pm 0.07$ h (\citealt{2018ATel11756....1P}) was identified from
intense photometric monitoring of the outburst phase.  

In this work we present the dynamical confirmation of the BH in J1820
employing time-resolved optical spectroscopy obtained while the source
was approaching its quiescent state. The letter is structured as
follows: Section \ref{sec:obs} presents the observations and the data reduction
steps. In Section \ref{sec:res} we establish the orbital ephemeris and
analyze the spectrum of the donor star.  Finally, in Section \ref{sec:disc} we discuss our results.

\section{Observations and data reduction} \label{sec:obs}

The observations of J1820 were carried out with the 10.4-m Gran Telescopio
Canarias (GTC) and 4.2-m William Herschel Telescope (WHT), both  at
the Observatorio del Roque de los Muchachos on La Palma, Spain.

We observed the target near the quiescent state in June 2019 
with the OSIRIS spectrograph
(\citealt{2000SPIE.4008..623C}) mounted on GTC. We used grism R2500R
combined with a 0\arcsec.6-wide slit and unbinned detector to cover the $5575-7685$ \AA~wavelength range with a 0.5 \AA
~pix$^{-1}$ dispersion and 2.5 \AA~FWHM spectral resolution (120 km~
s$^{-1}$ at H$\alpha$). In total we obtained 8
spectra across 4 different nights with integration times of 900~s
(see Table \ref{log}). Four 1200~s OSIRIS spectra of J1820 were 
also  obtained during a low
brightness state (\citealt{2019ATel12534....1R}) on the nights of 26 
and 27 Feb 2019 when the source brightness was  18.0 and 18.4-18.8, respectively,
according to the g-band acquisition images. The data were
taken with the R2500R grism, a 1\arcsec.0-wide slit and binned $2\times2$
detector, which delivered a spectral resolution of
4.4 \AA~(160 km~s$^{-1}$)~FWHM sampled with a 1.0~\AA~pix$^{-1}$.

Nine spectra were acquired on the night of 24 June 2019 UT with the
ISIS spectrograph  mounted on the WHT. The red arm of ISIS was used
with the R600R grating centered at 6150 ~\AA ~ and a 1\arcsec.0-wide slit to
observe the spectral range $5550-6750$~\AA~ with a
1.8~\AA~(80 km~s$^{-1}$)~FWHM resolution and a 0.5~\AA
~pix$^{-1}$ dispersion. The blue arm of the instrument was employed
with the R300B grating centered at  4500~\AA, covering
$3282-5300$~\AA~with a 4.1~\AA~FWHM resolution. The observations were split
into four visits equally spaced through the night to
sample close to half of the potential 17 h orbit. In each visit we
obtained $2\times1800$~s spectra, except 
on  the first visit when one extra 300~s spectrum was acquired.
The radial velocity standard stars of spectral types K5 V
(61 Cyg A) and K7 V (61 Cyg B) were taken during twilight using the same instrumental configuration as for J1820.

\begin{table}
\begin{center}
\caption{Journal of the J1820 observations.}
\begin{tabular}{lccccc}
\tableline\tableline
Date               & Instrument   & \#         & Exp.  & Res.  &   r  \\
(2019)              &              &      &   (s)      &
                                                                   (\AA)
                                                                         & (mag) \\
\\
26 Feb   &  OSIRIS    & 2   & 1200  &  3.4  & - \\
27 Feb   &  OSIRIS    & 2   & 1200  &  3.4  & - \\
7  Jun       & OSIRIS     & 1 &   900 &  2.5   &  17.5 \\
9 Jun       & OSIRIS     & 1 &   900 &  2.5   &  17.7-17.8\\

23 Jun        &  OSIRIS    &  3      &    900 &  2.5  & 17.5-17.7 \\
24 Jun        &  ISIS    & 9      &  300,1800 &  1.8  & -    \\
25 Jun       &  OSIRIS    &    3      &    900&    2.5 &  17.8-17.9 \\
\tableline
\end{tabular}
\end{center}
\label{log}
\end{table}

The spectra were reduced, extracted and wavelength calibrated using
standard techniques implemented in {\sc  iraf}.  Exposures of
comparison arc lamps were performed with OSIRIS after the end of each 
observing night while they were obtained 
bracketing the target observations with ISIS. The pixel-to-wavelength
scale was established through third order spline fits to HgAr+Ne and
CuNe+CuAr arc lines in the OSIRIS and ISIS lamp spectra, respectively. The
rms scatter of the fit was always $<0.01$ \AA~ (OSIRIS) and $<0.09$
\AA~ (ISIS). The [O{\sc i}]  6300.3 \AA~ sky line was used to  correct
for wavelength zeropoint deviations, which were $<30$ km s$^{-1}$
(OSIRIS) and $<10$ km s$^{-1}$  (ISIS). Each spectrum was
normalized by dividing it by 
a third-order spline fit 
to the continuum after masking out emission lines and telluric
bands. Finally, the spectra were rebinned onto a logarithmic wavelength scale. We used {\sc molly} and custom software under {\sc python} to
perform the analysis described in the next sections.

\subsection{Templates}
For the data analysis we also use a
set of high-resolution templates (R=55,000--110,000) from the spectral
libraries published by \citet{2004A&A...426..619E, 2006A&A...445..633E} and
\citet{1998A&AS..128..485M}. The name of the selected stars and
their spectral classification are given in Table \ref{tspcl}.

\section{Analysis and results}\label{sec:res}

\subsection{Radial velocity curve} \label{sec:rv}
Radial velocities were measured from the J1820 spectra by the method of
cross-correlation with a spectral template star (\citealt{1979AJ.....84.1511T}).
 On the basis of the results for the spectral classification presented below,
 we selected the spectrum of a K4 dwarf (HD 216803) for the cross-correlation. 
This was resampled onto the same logarithmic wavelength scale as for
the OSIRIS and ISIS spectra, and broadened to match the spectral resolution of these instruments
using the rotational profile of \citet{1992oasp.book.....G}.
For the analysis we selected the wavelength range $5200-6815$, common
to all data sets. This spectral  interval covers  several temperature and gravity sensitive
photospheric lines present in F to M stars that are useful for radial
velocity measurements as well as spectral-type and
luminosity classification. The radial velocities were extracted
from the 21 normalized individual J1820 spectra after masking interstellar
features and atmospheric bands present in the selected spectral interval. 

\begin{figure}
\includegraphics[scale=0.32,angle=-90]{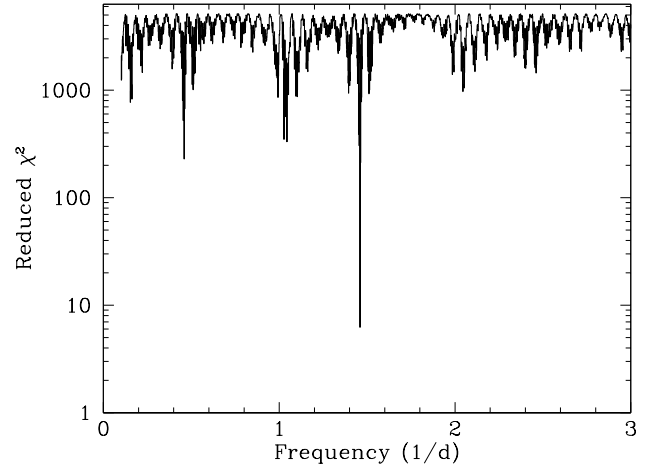}
\caption{$\chi^2$ periodogram of all our radial velocities.
The deepest peak is located at frequency 1.4588 d$^{-1}$ (=0.6855 d).
Note the logarithmic scale in the vertical axis.}
\label{fig1}
\end{figure}

\begin{figure}
\includegraphics[scale=0.42,angle=0]{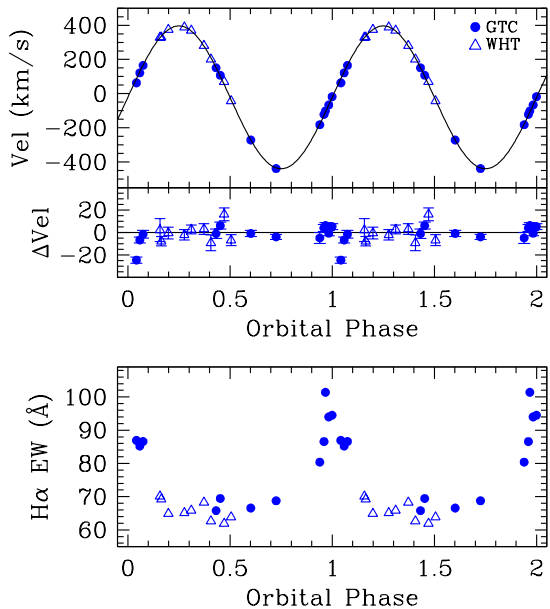}
\caption{Heliocentric radial velocities of the donor star (top) and
  H$\alpha$ EWs (bottom) phase-folded on the
ephemeris presented in section \ref{sec:rv}. The best sine wave fit to the radial velocities is 
overplotted. Residuals from the fit are shown in the middle panel. One orbital cycle is repeated for the sake of clarity. Note that error bars are smaller than data symbols.}
\label{fig2}
\end{figure}

The cross-correlation analysis of all our spectra reveals clear
radial velocity variations.  Thus, to determine an accurate orbital
period for J1820 we compute a $\chi^2$ periodogram of these velocities
in the frequency range 0.1 - 10 cycles d$^{-1}$ in 10$^{-5}$ frequency
steps (see Figure \ref{fig1}). The periodogram shows the lowest
minimum at 1.4588  cycles d$^{-1}$ (=0.6855 d). Other minima are
rejected as the potential orbital frequencies  give reduced $\chi{^2} > 200$.  
Least-squares sine fits to our radial velocity data
using the favoured periodicity as initial guess result in the following best fit parameters: 
\begin{itemize}
\item[] $T_{0} = HJD~2458540.043 \pm 0.002 $
\item[] $\gamma = -21.6  \pm 2.3$ km s$^{-1}$
\item[] $K_{2} = 417.7. \pm 3.9$ km s$^{-1}$
\item[] P$_{orb}$= 0.68549 $\pm$ 0.00001~d

\end{itemize}

where  $T_{0}$ is defined as the time of closest approach of the
donor star to the observer,  $\gamma$
is the heliocentric systemic velocity (corrected from the 
radial  velocity of the template used in the cross-correlation),  $K_{2}$ 
the  velocity semi-amplitude of
the donor star, and P$_{orb}$ 
the spectroscopic orbital
period. The four parameters were kept free during the fit.
 All quoted uncertainties are 1-$\sigma$.
These were obtained by scaling the error in the radial velocities
by a factor 1.7 to yield  a reduced $\chi{^2}=1$.  The phase-folded
radial velocity curve is shown in Figure \ref{fig2}.  

\begin{figure}
\includegraphics[scale=0.3,angle=0]{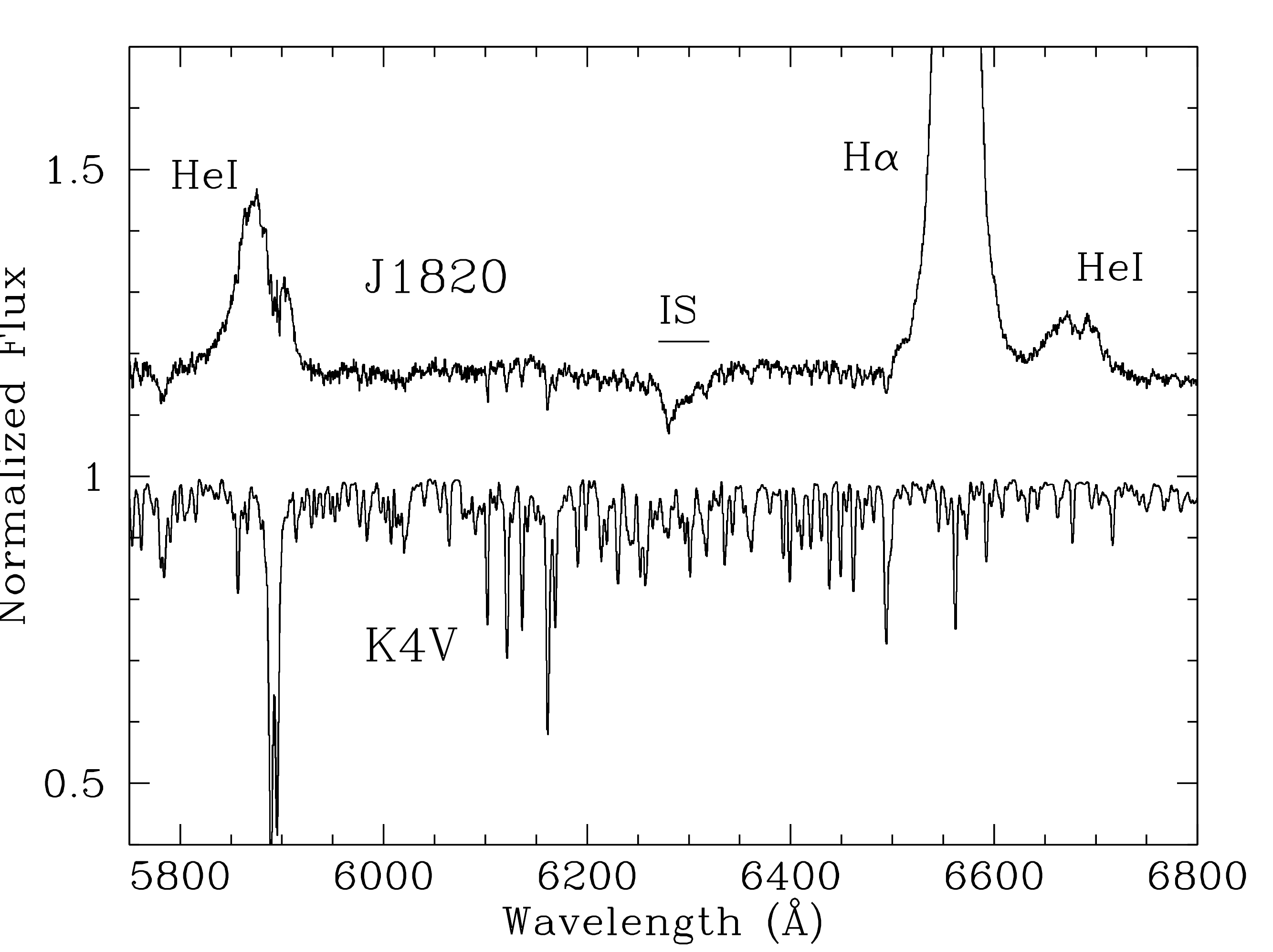}
\caption{Doppler-corrected average OSIRIS spectrum of J1820, 
compared with the best template spectrum broadened by 98 km s$^{-1}$. 
The spectra are normalized to the continuum and shifted vertically for
the sake of display.  The template contributes 18\% to the total light. 
The main spectral  features are identified. IS denotes an interstellar band contaminated
  with telluric 
  absorption.}
\label{fig3}
\end{figure}

\subsection{Spectral Type of the Donor Star}\label{sec:spcl}

To constrain the  spectral type of the donor star, we employ the technique
described in \citet{1994MNRAS.266..137M}, which consists on searching
for the lowest residual obtained after 
subtracting a set of normalized spectral templates from the average 
(Doppler-corrected) spectrum of the target. Prior to the subtraction, the
template spectra are broadened to match the 
width  of the absorption lines from the donor star.  This method allows
for the possibility of an extra light contribution to the total continuum
when searching for the  optimal  parameters
of the donor star.  We restricted this analysis to the J1820 
spectra obtained with OSIRIS
since they have the highest signal-to-noise ratio. 
Note here that the spectral resolution of the selected data is
  insufficient to establish with the described method the rotational broadening
  of the lines, while a reliable measurement of this broadening from
  the ISIS data is not  possible due to the different seeing-limited
  resolution of each spectrum imposed by variable 
 (0\arcsec.5-0\arcsec.9) image quality during the observations. As
templates we use the spectra of the two K-type dwarfs observed with ISIS and 
the set of high-resolution templates  described in section
\ref{sec:obs}. These stars have accurate stellar parameters ($\log g$, T$_{eff}$ and
metallicities) and, therefore, they 
provide a reference for a robust spectral-type classification. 

We implemented the optimal subtraction of the templates as follows:
first, the OSIRIS spectra were shifted to the rest frame of the donor star by
subtracting the radial velocities obtained from the cross-correlation
with the template. We averaged the individual  spectra 
with weights according to their signal-to-noise ratio in order to maximize the
signal-to-noise of their sum. On the other hand,
the template spectra were downgraded to the instrumental resolution of the OSIRIS data through convolution with the rotational profile of \citet{1992oasp.book.....G}. Each
broadened version of the template was multiplied by a factor $f$ (representing
the fractional contribution of light from the donor star to the 
continuum) and subtracted from the J1820 Doppler-corrected average. 
The spectral type and optimal value of $f$ were obtained 
by minimizing the $\chi^2$ between the residual of the subtraction and
a smoothed version of itself.  The results are listed in Table \ref{tspcl}. 
The minimization of  $\chi^2$ shows that the spectral type of the 
donor star in J1820 is most likely K3-5, 
while the low $f$ value indicates that the accretion disc contributes $\sim$80 \% to the total flux.
The  spectral classification should thus be considered preliminary given the strong dilution of the donor's absorption lines caused by the very large disc contribution.  
Figure \ref{fig3} presents the Doppler corrected OSIRIS spectrum of J1820, together with 
the best spectral template. 

In agreement with the large dilution factor, we note that 
the nightly r-band photometry measured
from the OSIRIS acquisition images 
shows evidence for significant flickering, with $\sim$0.2 mag amplitude, and 
a mean level $\sim1$ mag brighter than true quiescence (see
Table \ref{log}).  In addition, Figure \ref{fig2} reveals that the 
equivalent width (EW) of the H$\alpha$ line does not trace the classic orbital 
variation expected when the ellipsoidal modulation from the donor star is dominant 
(e.g. figure 6 in \citealt{1994MNRAS.266..137M}).    
Instead, we observe that the 
H$\alpha$ EW increases by a factor 1.5 when the
inner face of the donor star is hidden from view (orbital phase $\sim
0.0$). This behaviour may be the result 
of a grazing eclipse of the accretion disc by the 
donor star. Large EW jumps over a narrow orbital interval  
$\approx \pm 0.1 \times {\rm P_{orb}}$ centered at phase 0 are commonly seen 
in eclipsing cataclysmic variables (see e.g. \citealt{bianchini04}) 
and are explained as due to the fact that the emission lines are
formed in a chromosphere  above the accretion disc.  Alternatively, the sharp rise in EW at phase 0 could be explained by irradiation of the donor star if the
orbital inclination of the system is moderately high. However, we find
no evidence for spectral type variations induced by irradiation on the
Doppler-corrected averages of the GTC spectra obtained at orbital
phases 0.9-1.1 and 0.4-0.6. 
Neither we see 
evidence for irradiation effects in our radial velocity curve when examining the
residuals from the fit to the data (see middle panel in Figure 2). 
Therefore, we give more credit to the disc eclipse interpretation 
for the EW behaviour at phase 0. In either case, the change in EW
implies a drop in the r-band continuum of up to $\sim0.5 $ mag at
orbital phase 0.0 that should be detected in optical light curves. 

\begin{table}
\begin{center}
\label{spcl}
\caption{Spectral classification.}
\begin{tabular}{lcccc}
\tableline\tableline
Template     &Spectral     &$\chi^2$   &$f$       \\
 	            &Type	       &(d.o.f.=1087)    &           \\ 
\\
 
HD 39091   &   G0~V    &   1448.30 	      & 0.349   \\ 
HD 30495   &   G1.5~V &   1373.90 	      & 0.336  \\ 
HD 43162  &   G6.5~V  &   1316.5	      & 0.302   \\ 
HD 69830  &   K0~V     &   1260.2            & 0.291    \\ 
 HD 17925 &   K1.5~V  &   1188.84	      & 0.221   \\ 
HD 222237 &   K3~V   &   1132.96  	      & 0.214   \\ 
HD 216803 &    K4~V  &   1118.85	      & 0.176   \\ 
61 Cyg A$^a$     &    K5~V  &  1145.56     & 0.185   \\ 
61 Cyg B$^a$     &    K7~V  &   1200.54  & 0.173    \\ 
HD 157881 &    K7~V	   &   1171.48  & 0.151   \\ 
\tableline
\end{tabular}
\begin{tabular}{c}
\\
$^a$ Spectral templates obtained with ISIS.
\end{tabular}
\end{center}
\label{tspcl}
\end{table}     

\section{Discussion}\label{sec:disc}

Our values of $K_{2}$ and the orbital period yield a mass
function:
\noindent
\begin{displaymath} f(M) =
\frac{(M_1 \sin i)^3}{(M_1 + M_2)^2} = \frac{{P_{orb}}K_2^3}{2\pi G} = 5.18 \pm 0.15\
M_\sun, \end{displaymath}
\noindent
where $G$ is the gravitational constant and M$_1$ and M$_2$ are the compact
object and donor star mass, respectively. Given that the mass function
provides a lower limit on the mass of the compact object and this
exceeds in J1820 the maximum mass allowed for a neutron star ($\sim3$
M$_\odot$), we conclude that J1820 is a BH. The donor star in J1820 is constrained to be K3-K5. A K3 V donor will fill its Roche lobe  when the binary orbital period
is P$_{orb}$ (hr) $\approx \sqrt{110/\bar{\rho} (g cm^{-3})}$ = 6.4
(\citealt{2002apa..book.....F}), where  we have used a K3 dwarf  mean density 
$\bar{\rho}\approx 2.7$ g cm$^{-3}$
(\citealt{2012ApJ...757..112B}). Given the orbital period of J1820,
the donor star must  clearly  be evolved in order to fill its Roche lobe.
The presence of evolved donor stars in accreting binaries is not
uncommon. For instance, this is the case for the BH X-ray transient Nova Ophiuchi 1977
(\citealt{1997AJ....114.1170H}) that, with a 0.522 d orbital period, is a close match to J1820.

The $0.68549  \pm 0.00001$ d spectroscopic orbital period is 2.6\% shorter than
the photometric period measured during outburst and therefore
consistent with the latter not being the true orbital period but a
superhump periodicity ($P_{sh}$) caused by a precessing disc
(e.g. \citealt{1996MNRAS.282..191O}, \citealt{2002MNRAS.333..791Z}). 
The  superhump is  the beat  frequency between the orbital and disc
precession frequencies: $P_{\rm  prec} =  \left(P_{\rm  orb}^{-1} -  P_{\rm
sh}^{-1} \right)^{-1} $. This implies $P_{\rm  prec} \approx 28
$ d during outburst. Moreover, the binary mass ratio $q={M_2}/{M_1}$ can
be estimated using the relation  between the period excess
$\Delta P = (P_{sh} - P_{orb})/ P_{orb}$ and $q$. For $q$ in the range
0.04--0.30, $\Delta P \simeq 0.216(\pm0.018)\,q$ (\citealt{2001PASP..113..736P,2005PASP..117.1204P}). Thus we  derive
$q\simeq 0.12$ for J1820.  The shape of the hardness-intensity diagram (\citealt{2019arXiv190604835M,2013MNRAS.432.1330M}), the detection of X-ray dips during
the hard state (\citealt{2018ATel11576....1H, 2019arXiv190606519K}) and the 
behaviour of the H$\alpha$ EW at phase 0 indicate that the orbital inclination 
must be moderately high. Given that we likely detected a grazing eclipse in the H$\alpha$ EW, the constraint on $q$ 
can be used  to set a lower limit to the orbital inclination, 
using the disc eclipse condition through e.g. equation 5 in \cite{casares18}. Assuming a typical disc 
size $R_{\rm d}\sim0.5~R_{\rm L1}$, where $R_{\rm L1}$ stands for the equivalent radius of the 
compact object's Roche lobe, we obtain  $i\gtrsim69^{\circ}$. 
An upper limit to the inclination is provided by the non detection of
X-ray eclipses during outburst through $\cos i \ge 0.462 \left( q/ 1+q
\right)^{1/3}$ i.e. $i\lesssim77^{\circ}$. These constraints to the
binary inclination imply a BH mass M${_1} = 7.0-8.0$ M$_\odot$,  although we 
caution that this calculation assumes $q\sim0.12$, based on a scaling relation of 
the superhump period excess, and requires confirmation.
For example, if the mass ratio were $q=0.07$ the inclination limits would be 
$i\sim71^{\circ}-79^{\circ}$ and the BH mass M${_1} \sim 6.3-6.9$ M$_\odot$. 

Interestingly, our accurate ephemeris allows us also to compute the
orbital phase $\phi $ of the $XMM-Newton$ 
dipping episode reported in \cite{2019arXiv190606519K}. 
We obtain $\phi_{\rm dip}=0.87\pm0.01$, thus confirming the X-ray dip was caused by 
absorption in the disc bulge. We therefore conclude that J1820 is another high-inclination BH dipper, analogous to GRO J1655-40, 4U 1630-47 and MAXI J1659-152 \citep{kuulkers98, kuulkers13}.  

\citet{2019arXiv190604835M} presented the discovery of an accretion
disc outflow during the hard state of the outburst of J1820. This
result, together with previous findings on the long-orbital period
($>2$ d) BH transients V404 Cyg and V4641 Sgr
(\citealt{2016Natur.534...75M},\citealt{2018MNRAS.479.3987M}),
shows that winds are likely a common feature of BH transients in
outburst.  
Our determination of a 
0.68 d orbital period for J1820, 
however, seems to indicate that the presence of winds is not limited to the BH
transients with the largest accretion discs. 

With J1820, there are now 19 Galactic X-ray transients that
have been dynamically confirmed to contain a BH.
Further progress relies on a dynamical study during true
quiescence, when the disc contribution to the total light will have
decreased. In particular, higher resolution spectra will permit an
accurate evaluation of the rotational broadening and hence the system mass ratio. The orbital inclination may also be further constrained by modelling the ellipsoidal modulation of the companion star at optical and infrared wavelengths. 
It is interesting to note that our measurement of a systemic velocity 
$\gamma=-22\pm2$ km s$^{-1}$, together with the Gaia DR2 proper motion and distance 
determinations, implies a large peculiar velocity $\sim$100 km s$^{-1}$ for J1820 
(see Figure 8 in \citealt{2019MNRAS.485.2642G}). A future accurate BH mass determination in J1820 will be important to 
continue building a potential anti-correlation between BH masses and kick velocities and, thus, 
help constrain BH formation channels. 

\section{acknowledgements}
We are greatful  to the GTC and ING staff, in particular Antonio
L. Cabera Lavers and Ian Skillen, for their help to implement the
observations presented in this work.  We thank the referee for
  useful comments.
We acknowledge support by the Spanish MINECO under grant
AYA2017-83216-P. TMD and MAPT acknowledge support via Ram\'on y Cajal
Fellowships RYC-2015-18148 and RYC-2015-17854.  
PGJ acknowledges funding from the European Research Council under ERC
Consolidator Grant agreement no 647208. {\sc molly} software developed
by Tom Marsh is gratefully acknowledged. {\sc iraf} is distributed by
the National Optical Astronomy Observatory, which is operated by the
Association of Universities for Research in Astronomy (AURA) under a
cooperative agreement with the National Science Foundation. 
We dedicate this letter to the memory of Jeff E. McClintock.



\end{document}